\documentstyle[aps]{revtex}
\newcommand{\bb}{\begin{equation}}
\newcommand{\ee}{\end{equation}}
\newcommand{\ba}{\begin{array}}
\newcommand{\ea}{\end{array}}

\begin {document}

\title{Electron-positron pair creation in a vacuum by an
        electromagnetic field in 3+1 and lower dimensions}
\author{Qiong-gui Lin\thanks{E-mail addresses: qg\_lin@163.net,
stdp@zsu.edu.cn}}
\address{China Center of Advanced Science and Technology (World
	Laboratory),\\
        P.O.Box 8730, Beijing 100080, People's Republic of China\\
        and\\
        Department of Physics, Zhongshan University, Guangzhou
        510275,\\
        People's  Republic of China \thanks{Mailing address}}

\maketitle
\vskip 0.3cm
\begin{abstract}
{\normalsize
We calculate the probability of electron-positron pair creation
in vacuum in 3+1 dimensions by an external electromagnetic
field composed of a constant uniform electric field and a
constant uniform magnetic field, both of arbitrary magnitudes
and directions. The same problem is also studied in 2+1 and
1+1 dimensions in appropriate external fields and similar
results are obtained.}
\end{abstract}
\vskip 0.3 cm
\leftline {PACS number(s): 03.70.+k, 11.15.Tk, 11.10.Kk}

\section{Introduction}               

Pair creation of charged particles in vacuum by an external
electric field was first studied by Schwinger several
decades ago [1]. Related problems have been discussed by
many authors, for example, in Refs. [2-6]. Though half
a century has passed since the publication of Schwinger's
classic work, the subject of pair creation remains a
widely discussed one in string theory and black hole
theory in these days.

It seems that most authors of the cited works dealt with
the problem only in electric fields. A magnetic field,
on the other hand, was left out in the cold. This may be
due to the fact that a pure magnetic field does not
lead to creation of particle-antiparticle pairs. In spite
of this fact, the presence of a magnetic field does change
the probability of pair creation by a pure electric field.
This can be easily seen when the magnetic field {\bf B} is
perpendicular to the electric field {\bf E}. The situation
may be changed by a Lorentz boost to another one with a pure
electric field, provided $|{\bf E}|>|{\bf B}|$. The result is
easily obtained, and is obviously different from the case
where {\bf B} is absent. When {\bf B} is not perpendicular
to {\bf E}, the problem is more complicated. The result for
the general case cannot be easily obtained from previous ones
and needs further study. As an exact and nonperturbative result
can be achieved as in the case of a pure electric field, the
study is worthwhile and may be of some interest.

In this paper we consider a magnetic field {\bf B} as well as
an electric field {\bf E}, both being constant and uniform, but
with arbitrary magnitudes and directions. The field of electrons
(or other charged fermions of spin $1 \over 2$) is second
quantized, while the electromagnetic field is treated classically
as a background field for the electrons. The probability of
electron-positron pair creation in vacuum is calculated exactly.
More specifically, we first deal with the relatively simple  case
where ${\bf B}\parallel {\bf E}$ ({\bf B} points at the same or
opposite direction of {\bf E}). For the general case, one can
always find by making a Lorentz boost an inertial frame $K'$
where the transformed fields satisfy ${\bf B}'\parallel {\bf E'}$.
In the system $K'$ the result is obtained in terms of ${\bf E'}$
and ${\bf B'}$. As the probability should be invariant under
Lorentz  boost, we obtain the result in the original system by
using the relation between $({\bf E', B'})$ and $({\bf E, B})$.
When ${\bf E}=0$ the probability vanishes while for ${\bf B}=0$
it reduces to Schwinger's result, as expected. This is done
in Sec. II.

In previous works main attention was paid to the problem in
ordinary 3+1 dimensions. In lower spatial dimensions the result
might be expected to be somewhat different. In Sec. III we turn
to the problem in 2+1 dimensions. In this case the magnetic
field has only one component and the electric field has two,
thus the problem is simpler than in 3+1 dimensions. In Sec. IV
we calculate the result in 1+1 dimensions. In this case there is no
magnetic field and the electric field has only one component, so
the problem is still simpler. The results in lower dimensions
are similar to that in 3+1 dimensions, but cannot be trivially
obtained from the latter. The difference among them can be
easily seen from the appearances of the corresponding results.
These are summarized in Sec. V.

\section{3+1 dimensions}          

We use natural units where $\hbar=c=1$ throughout this paper.
Consider the electron with mass $m$ and charge $e<0$ (the
following results are applicable to other charged fermions
of spin $1 \over 2$), moving in a background electromagnetic
field described by the vector potential $A_\mu$. The field of the
electron is second quantized, while $A_\mu$ is treated classically.
The vacuum-vacuum transition amplitude can be shown to be [1, 2]
\bb
S_0=\exp\left[-{\rm Tr}\ln{\gamma\cdot P-m+i\epsilon \over
\gamma\cdot(P-eA)-m+i\epsilon}\right], 
\ee            
where $\epsilon = 0^+$, and the Tr indicates a complete diagonal
summation over the space-time coordinates as well as the spinorial
indices. In this section we consider the problem in
ordinary 3+1 dimensions. But Eq. (1) holds in 2+1 and 1+1 dimensions
as well. In Eq. (1) $X_\mu$ (the independent variables of $A_\mu$)
and $P_\mu$ are now operators satisfying
\bb
X_\mu|x\rangle=x_\mu|x\rangle,\quad
\langle x|P_\mu|\varphi\rangle=i\partial_\mu
\langle x|\varphi\rangle,
\ee     
where $|\varphi\rangle$ is an arbitrary state. Consequently
\bb
[X_\mu, P_\mu]=-ig_{\mu\nu},
\ee     
where $g_{\mu\nu}={\rm diag}(1,-1,-1,-1)$ and similarly in lower
dimensions. Using the charge conjugation matrix and the fact that
the trace of an operator is invariant under matrix transposition
we have
\bb
S_0=\exp\left[-{\rm Tr}\ln{\gamma\cdot P+m-i\epsilon \over
\gamma\cdot(P-eA)+m-i\epsilon}\right].
\ee            
This holds in lower dimensions as well. Multiplying Eqs. (1) and
(4) we have
\bb
S_0^2=\exp\left\{-{\rm Tr}\ln{P^2-m^2+i\epsilon \over
[\gamma\cdot(P-eA)]^2-m^2+i\epsilon}\right\}.
\ee            
Taking the module we obtain the vacuum-vacuum transition probability
in the form
\bb
|S_0|^2=\exp\left[-\int dx\; w(x)\right],
\ee     
where
\bb
w(x)={\rm Re}\left\langle x\left|{\rm tr}\ln{P^2-m^2+i\epsilon
\over [\gamma\cdot(P-eA)]^2-m^2+i\epsilon}\right|x\right\rangle
\equiv{\rm Re}W(x)
\ee            
is to be interpreted as the probability, per unit time and per unit
volume, at the space-time position $x$, of electron-positron
pair creation by the external electromagnetic field. For constant
uniform electromagnetic field it is expected to be independent of
$x$. In the above equation the tr indicates ordinary diagonal
summation over spinorial indices. Using the identity
\bb
\ln{a+i\epsilon\over b+i\epsilon}=\int_0^\infty\;{ds\over s}
[e^{is(b+i\epsilon)}-e^{is(a+i\epsilon)}]
\ee     
and the ralation
\bb
[\gamma\cdot(P-eA)]^2=(P-eA)^2-{e\over 2}\sigma^{\mu\nu}
F_{\mu\nu}
\ee     
where
\bb
F_{\mu\nu}=\partial_\mu A_\nu-\partial_\nu A_\mu, \quad
\sigma^{\mu\nu}={i\over 2}[\gamma^\mu,\gamma^\nu],
\ee     
we have
\bb
\ln{P^2-m^2+i\epsilon\over [\gamma\cdot(P-eA)]^2-m^2+i\epsilon}
=\int_0^\infty\;{ds \over s}e^{-is(m^2-i\epsilon)}\left[e^{is(P-eA)^2}
\exp\left(-{i\over2}es\sigma^{\mu\nu}F_{\mu\nu}\right)-e^{isP^2}
\right].
\ee     
Up to this point we have only reviewed some results obtained by
previous authors [1,2]. These are necessary for further discussions.
We emphasize that Eqs. (7) and (11) are valid in lower dimensions
as well as in 3+1 dimensions. The following calculations depend on
the spatial dimension, thus we will henceforth deal only with the
(3+1)-dimensional case in this section. We will return to the
lower-dimensional case in the following sections.

Now we consider a constant uniform electromagnetic field where
${\bf B\parallel E}$. Without loss of generality we take
\bb
{\bf E}=E{\bf e}_x,\quad {\bf B}=B{\bf e}_x,
\ee     
where ${\bf e}_x$ is the unit vector in the $x^1$ direction, $E$ and
$B$ are constants which may be positive or negative. We have then
for $F_{\mu\nu}$ the nonvanishing components $F_{01}=E$, $F_{23}=-B$,
and
$$
-\textstyle{1\over 2}i\sigma^{\mu\nu}F_{\mu\nu}
=\gamma^0\gamma^1 E-\gamma^2\gamma^3 B.
$$
Using the properties of the $\gamma$ matrices, we have
\bb
\left(-\textstyle{1\over 2}i\sigma^{\mu\nu}F_{\mu\nu}\right)^2
=E^2-B^2+i2EB\gamma_5,
\ee                         
where $\gamma_5=\gamma^5=i\gamma^0\gamma^1\gamma^2\gamma^3$. As
$\gamma_5^2=1$, the eigenvalues of $\gamma_5$ are $\pm 1$, both being
double degenerate. Thus the eigenvalues of $(-i\sigma^{\mu\nu}F_
{\mu\nu}/2)^2$ are $(E\pm iB)^2$, both being double degenerate,
and the four eigenvalues of $-i\sigma^{\mu\nu}F_{\mu\nu}/2$
are $E\pm iB$, $-(E\pm iB)$. Therefore we have
\bb
{\rm tr}\,\exp\left(-\textstyle{\frac 12}ies\sigma^{\mu\nu}
F_{\mu\nu}\right)=4\cosh(eEs)\cos(eBs),
\ee         
and
\bb
{\rm tr}\ln{P^2-m^2+i\epsilon\over [\gamma\cdot(P-eA)]^2
-m^2+i\epsilon}=4\int_0^\infty\;{ds \over s}e^{-is(m^2-i\epsilon)}
[\cosh(eEs)\cos(eBs)e^{is(P-eA)^2}-e^{isP^2}].
\ee     
The next step is to calculate the matrix elements
$\langle x|e^{is(P-eA)^2}|x\rangle$ and $\langle x|e^{isP^2}|x
\rangle$. The second is easy. We have
\bb
\langle x|e^{isP^2}|x\rangle
=\int dk\; \langle x|e^{isP^2}|k\rangle\langle k|x\rangle,
\ee     
where
\bb
\langle x|k\rangle={1\over (2\pi)^2}e^{-ik\cdot x},\quad
\langle k|x\rangle={1\over (2\pi)^2}e^{ik\cdot x}.
\ee     
Using Eq. (2) it is easy to show that
\bb
P_\mu|k\rangle=k_\mu|k\rangle,\quad
\langle k|P_\mu=k_\mu\langle k|.
\ee     
Substituting Eqs. (17) and (18) into Eq. (16) we obtain
\bb
\langle x|e^{isP^2}|x\rangle
=\int dk\; e^{isk^2}\langle x|k\rangle\langle k|x\rangle
={1 \over (2\pi)^4}\int dk\;e^{isk^2}=-{i\over 16\pi^2s^2}.
\ee     
The calculation of another matrix element is somewhat complicated.
We denote
\bb
X^\mu=(X_0, X,Y,Z),\quad P^\mu=(P_0,P_x,P_y,P_z),
\ee              
and choose
\bb
A^\mu=(A_0,A_x,A_y,A_z)=(-EX,0,0,BY),
\ee                 
which corresponds to the field strengths (12), then
\bb
(P-eA)^2=(P_0+eEX)^2-(P_z-eBY)^2-P_x^2-P_y^2.
\ee        
It is not difficult to show that
$$
(P_0+eEX)^2=e^{iP_0P_x/eE}e^2E^2X^2e^{-iP_0P_x/eE},
\eqno(23{\rm a})$$
$$
(P_z-eBY)^2=e^{-iP_yP_z/eB}e^2B^2Y^2e^{iP_yP_z/eB}.
\eqno(23{\rm b})$$
\addtocounter{equation}{1}
Substituting these ralations into Eq. (22) we have
\bb
(P-eA)^2=-e^{iP_0P_x/eE}(P_x^2-e^2E^2X^2)e^{-iP_0P_x/eE}-
e^{-iP_yP_z/eB}(P_y^2+e^2B^2Y^2)e^{iP_yP_z/eB},
\ee     
and thus
\bb
e^{is(P-eA)^2}=e^{iP_0P_x/eE-iP_yP_z/eB}e^{-is(P_x^2-e^2E^2X^2)}
e^{-is(P_y^2+e^2B^2Y^2)}e^{-iP_0P_x/eE+iP_yP_z/eB}.
\ee                
With these preparations we write down
\bb
\langle x|e^{is(P-eA)^2}|x\rangle=\int dk\;dk'\;\langle x|k'\rangle
\langle k'|e^{is(P-eA)^2}|k\rangle\langle k|x\rangle.
\ee     
Using Eqs. (25), (17), (18), and $\delta[(k_x'-k_x)/eE]=|eE|
\delta(k_x'-k_x)$ etc., after some algebras we arrive at
\bb
\langle x|e^{is(P-eA)^2}|x\rangle={|eE||eB|\over 4\pi^2}
{\rm tr}\,e^{-is(P_x^2-e^2E^2X^2)}{\rm tr}\,e^{-is(P_y^2+e^2B^2Y^2)},
\ee        
where
$$
{\rm tr}\,e^{-is(P_x^2-e^2E^2X^2)}=\int dk_x\;\langle k_x|
e^{-is(P_x^2-e^2E^2X^2)}|k_x\rangle
$$
and similarly for another trace, where $k_x$ is the first spatial
component of $k^\mu$. By comparison with the harmonic oscillator
or by using the technique of path integral, one can find the following
results.
$$
{\rm tr}\,e^{-is(P_x^2-e^2E^2X^2)}={1\over 2\sinh(|eE|s)},
\eqno(28{\rm a})$$
$$
{\rm tr}\,e^{-is(P_y^2+e^2B^2Y^2)}=-{i\over 2\sin(|eB|s)}.
\eqno(28{\rm b})$$
\addtocounter{equation}{1}
Substituting these into Eq. (27) we obtain
\bb
\langle x|e^{is(P-eA)^2}|x\rangle=-{i|eE||eB|\over 16\pi^2
\sinh(|eE|s)\sin(|eB|s)}.
\ee     
Combining the results (7), (15), (19), and (29) we arrive at
\bb
W(x)={1\over 4\pi^2 i}\int_0^\infty ds\;{e^{-im^2s}\over s}\left[
|eE||eB|\coth(|eE|s)\cot(|eB|s)-{1\over s^2}\right],
\ee            
where the $\epsilon$-dependent term in the exponential has been
dropped as it is no longer necessary. This is obviously independent
of $x$ as expected. Using the identity ${\rm Re}W=(W+W^*)/2$ and
making the change of variable $s\to -s$ in $W$ we obtain
\bb
w=-{1\over 8\pi^2 i}\int_{-\infty}^{+\infty} ds\;{e^{im^2s}\over s}
\left[|eE||eB|\coth(|eE|s)\cot(|eB|s)-{1\over s^2}\right].
\ee            
It is easy to see that the integrand has singularities (simple
poles) at $0$ and
$\pm n\pi/|eB|$ ($n=1,2,\ldots$) in the integration path. Thus the
integral in the above equation is not well defined. An appropriate
prescription should be employed. There are mainly three different
prescriptions. The first is to replace the integration path by a
straight line a bit above the real axis on the complex $s$ plane,
i.e., a straight line from $-\infty+i\epsilon$ to $+\infty+i\epsilon$
where $\epsilon=0^+$. The second is to replace the integration
path by one from $-\infty-i\epsilon$ to $+\infty-i\epsilon$. The
third is to keep the original path but take the Cauchy principal
value. It turns out that only the first prescription gives a
physically acceptable result. The second or the third will give a
negative result for $w$ when $B=0$, and thus are not acceptable.
When $B=0$ the integrand has one singularity in the integration
path, the simple pole at $s=0$. An appropriate prescription is also
needed in this case. It is the first prescription described above
that was implicitly used in Ref. [2]. The validity of the prescription
was confirmed by the coincidence of the result with
that of Schwinger obtained by a different method. We adopt this
prescription, and close the contour of integration at infinity
by a semicircle in the upper half plane. The integrand has simple
poles at $n\pi i/|eE|$ ($n=1,2,\ldots$) in the upper half plane.
The integral can be evaluated by using the residue theorem,
and the result turns out to be
\bb
w_\parallel={e^2|EB|\over 4\pi^2}\sum_{n=1}^\infty{1\over n}\coth
\left(n\pi{|B|\over |E|}\right)\exp\left(-{n\pi m^2\over|eE|}\right).
\ee                   
This is the result for the simple case ${\bf B\parallel E}$, as
indicated by the subscript. The convergence of the series is obvious.
It is easy the see that $w=0$ when $E=0$, which means that a pure
magnetic field cannot lead to pair creation as expected. For $B=0$,
on the other hand, we use $\coth u\to 1/u$ ($u\to 0$), and obtain
\bb
w_E={e^2E^2\over 4\pi^3}\sum_{n=1}^\infty{1\over n^2}
\exp\left(-{n\pi m^2\over |eE|}\right),
\ee                   
where the subscript indicates a pure electric field. This is just
Schwinger's result. When $|B|=|E|$, we have
\bb
w_{|B|=|E|}={e^2E^2\over 4\pi^2}\sum_{n=1}^\infty{1\over n}\coth
(n\pi)\exp\left(-{n\pi m^2\over |eE|}\right).
\ee                   
Since $\coth(n\pi)>1$ and $1/n\ge1/n^2$, the series in Eq. (34) is
obviously larger than that in Eq. (33), and the overall factor in
Eq. (34) is also larger than the one in Eq. (33) by a factor $\pi$,
hence $w_{|B|=|E|}>\pi w_E$. However, since $|eE|\ll m^2$ as currently
available, the dominant term in both Eqs. (33) and (34) is the first
term, and thus $w_{|B|=|E|}\approx\pi w_E$. If $B$ increases, the
result may be still larger. However, If $E$ cannot be significantly
raised, the addition of a magnetic field cannot raise the probability
greatly.

We can now turn to the general case where both {\bf E} and {\bf B}
have arbitrary magnitudes and directions. We can always find another
frame of reference $K'$ by a Lorentz boost where the transformed
fields ${\bf E'}$ and ${\bf B'}$ satisfy ${\bf B'\parallel E'}$.
In the system $K'$ we can find the probability in terms of ${\bf E'}$
and ${\bf B'}$ by using Eq. (32).
We know that the probability is a Lorentz
scalar. This is because the total probability $\int dx\; w(x)$ is a
Lorentz scalar as it only involves a process of number counting, and
the space-time volume $dx$ is also invariant under Lorentz boost.
Hence  the probability in the original system can be obtained by
using the relation between ({\bf E, B}) and (${\bf E', B'}$). Another
approach to the result is also available. Since $w(x)$ is a Lorentz
scalar, it can only involve Lorentz invariants constructed from
{\bf E} and {\bf B}. In the special case ${\bf B\parallel E}$, it
must reduces to the result (32). This enables us to obtain the
general result more easily. We define the Lorentz invariants
$$
{\cal F}={\bf E^2-B^2},\quad {\cal G}=2{\bf E\cdot B},
\eqno(35{\rm a})$$
$$
{\cal E}=\left({\sqrt{{\cal F}^2+{\cal G}^2}+{\cal F}\over 2}
\right)^{1\over2},
\quad
{\cal B}=\left({\sqrt{{\cal F}^2+{\cal G}^2}-{\cal F}\over 2}
\right)^{1\over2},
\eqno(35{\rm b})$$
\addtocounter{equation}{1}
then the result reads
\bb
w={e^2{\cal EB}\over 4\pi^2}\sum_{n=1}^\infty{1\over n}\coth
\left(n\pi{\cal B \over E}\right)\exp\left(-{n\pi m^2\over|e|{\cal E}}
\right).
\ee                   
Now we can examine the result by another special case where
${\bf B\perp E}$. In this case we have ${\cal G}=0$, and
$$
{\cal E}=\sqrt{\bf E^2-B^2},\quad {\cal B}=0
\eqno(37{\rm a})$$
if $|{\bf E}|>|{\bf B}|$, or
$$
{\cal E}=0,\quad {\cal B}=\sqrt{\bf B^2-E^2}
\eqno(37{\rm b})$$
\addtocounter{equation}{1}
if $|{\bf E}|<|{\bf B}|$. The case (37a) is equivalent to that of a
pure electric field and the result is given by Eq. (33) where $|E|$
is replaced by ${\cal E}=\sqrt{\bf E^2-B^2}$. The case (37b) is
equivalent to that of a pure magnetic field and $w=0$. These are
all expected results and further confirm the result (36).

\section{2+1 dimensions}        

In this section we calculate the probability of pair creation in
vacuum in 2+1 dimensions. This cannot be trivially obtained from the
result in 3+1 dimensions. In two spatial dimensions the magnetic field
has only one component, while the electric field has two. There is
nothing like ${\bf E\cdot B}$. The only Lorentz invariant constructed
from {\bf E} and $B$ is ${\bf E}^2-B^2$. Consider an electromagnetic
field ({\bf E}, $B$) where both {\bf E} and $B$ are constant and
uniform. There are two different situations to be distinguished. The
first one is characterized by the inequality $|{\bf E}|>|B|$, while
the second by $|{\bf E}|<|B|$. The first situation is equivalent to
one with a pure electric field. One can calculate the result for the
simple case with a pure electric field, and get the result for the
general case by the method of Sec. II. This will be discussed in
detail in the following. The second situation is equivalent to one
with a pure magnetic field. Similar calculations give a vanishing
probability in this case. This is an expected result and we will not
discuss it in detail.

In Sec. II we have emphasized that Eqs. (7) and (11) are valid in
lower dimensions. We will begin with these equations. Consider a
pure electric field {\bf E} which is constant and uniform. Without
loss of generality we choose
\bb
{\bf E}=E{\bf e}_x,
\ee            
where $E$ is a constant which may be positive or negative. We have
then for $F_{\mu\nu}$ the nonvanishing component $F_{01}=E$, and
\bb
\exp\left(-\textstyle{\frac 12}ies\sigma^{\mu\nu}
F_{\mu\nu}\right)
=\exp(eEs\gamma^0\gamma^1).
\ee         
As $(\gamma^0\gamma^1)^2=1$, and ${\rm tr}(\gamma^0\gamma^1)=0$,
the trace of the above expression can be evaluated directly with
the result
\bb
{\rm tr}\,\exp\left(-\textstyle{\frac 12}ies\sigma^{\mu\nu}
F_{\mu\nu}\right)=2\cosh(eEs).
\ee         
Note that in 2+1 dimensions the $\gamma$ matrices are $2\times 2$
ones and thus ${\rm tr}\,{\bf 1}=2$.
Combining Eqs. (40) and (11) we have
\bb
{\rm tr}\ln{P^2-m^2+i\epsilon\over [\gamma\cdot(P-eA)]^2
-m^2+i\epsilon}=2\int_0^\infty\;{ds \over s}e^{-is(m^2-i\epsilon)}
[\cosh(eEs)e^{is(P-eA)^2}-e^{isP^2}].
\ee     
The next step is to evaluate the matrix elements
$\langle x|e^{is(P-eA)^2}|x\rangle$ and $\langle x|e^{isP^2}|x
\rangle$. The second one can be easily worked out with the result
\bb
\langle x|e^{isP^2}|x\rangle
={1-i\over 4(2\pi)^{3\over 2}s^{3\over 2}}.
\ee     
This is rather different from the corresponding result in 3+1
dimensions, but the calculation is similar. For the first one,
we denote
\bb
X^\mu=(X_0, X,Y),\quad P^\mu=(P_0,P_x,P_y),
\ee              
and choose
\bb
A^\mu=(A_0,A_x,A_y)=(-EX,0,0),
\ee                 
which corresponds to the field strength (38), and results in
\bb
(P-eA)^2=(P_0+eEX)^2-P_x^2-P_y^2.
\ee        
Using Eq. (23a) we have
\bb
e^{is(P-eA)^2}=e^{iP_0P_x/eE}e^{-is(P_x^2-e^2E^2X^2)}
e^{-iP_0P_x/eE}e^{-isP_y^2}.
\ee                
The subsequent calculations are similar to those carried out in
obtaining Eq. (27) but simpler, the result reads
\bb
\langle x|e^{is(P-eA)^2}|x\rangle={(1-i)|eE|\over 2(2\pi)^{3\over2}
\sqrt s}
{\rm tr}\,e^{-is(P_x^2-e^2E^2X^2)}.
\ee        
Using Eq. (28a) we obtain
\bb
\langle x|e^{is(P-eA)^2}|x\rangle={(1-i)|eE|\over 4(2\pi)^{3\over2}
\sqrt s \sinh(|eE|s)}.
\ee     
Combining Eqs. (7), (41), (42), and (48) we arrive at
\bb
W(x)={1-i\over 2(2\pi)^{3\over2}}\int_0^\infty ds\;
{e^{-im^2s}\over s^{3\over2}}\left[
|eE|\coth(|eE|s)-{1\over s}\right].
\ee            
If we are going to use the residue theorem for contour integrals, we
must treat the integrand carefully since it is a multivalued function
in the complex $s$ plane. We cut the $s$ plane from 0 to $-i\infty$
along the imaginary axis, and define $\arg s=0$ in the positive
real axis, then the integrand is single valued in the cut plane.
We use the identity ${\rm Re}W=(W+W^*)/2$ and
making the change of variable $s\to s'=e^{i\pi}s$ in $W$ to yield
\bb
w={1+i\over 4(2\pi)^{3\over2}}\int_{-\infty}^{+\infty} ds\;
{e^{im^2s}\over s^{3\over2}}\left[
|eE|\coth(|eE|s)-{1\over s}\right].
\ee            
Now the integrand has one singularity in the integration path, the
origin $s=0$. This is a branch point of the integrand. Moreover, the
integrand tends to infinity like $1/\sqrt s$ when $s\to 0$. As the
lower half plane has been cut, we have now two different prescriptions
for the integral: to replace the integration path by one from
$-\infty+i\epsilon$ to $+\infty+i\epsilon$, or to take the Cauchy
principal value. It turns out that the two prescriptions give the same
result. We adopt the first prescription, close the contour of
integration at infinity by a semicircle in the upper half plane and
use the residue theorem to evaluate the integral. Note that there are
simple poles $n\pi e^{i\pi/2}/|eE|$ ($n=1,2,\ldots$) of the integrand
in the upper half plane. The result turns out to be
\bb
w_E={|eE|^{3\over2}\over 4\pi^2}\sum_{n=1}^\infty{1\over n^{3\over2}}
\exp\left(-{n\pi m^2\over |eE|}\right),
\ee                   
where the subscript indicates a pure electric field. 

The case of a pure magnetic field $B$ can be treated in a similar way.
It turns out that the probability vanishes as expected.

To conclude this section we consider the general case of an
electromagnetic field ({\bf E}, $B$) where {\bf E} has an arbitrary
direction. As pointed out at the beginning of this section, one must
distinguish between the two different cases $|{\bf E}|>|B|$ and
$|{\bf E}|<|B|$. The results can be easily obtained by the method
of Sec. II. When $|{\bf E}|>|B|$ we have
$$
w={|e|^{3\over2}{\cal E}^{3\over2} \over 4\pi^2}\sum_{n=1}^\infty
{1\over n^{3\over2}}
\exp\left(-{n\pi m^2\over |e|{\cal E}}\right),
\eqno(52{\rm a})$$
where
$$
{\cal E}=\sqrt{{\bf E}^2-B^2}.
\eqno(52{\rm b})$$
\addtocounter{equation}{1}
When $|{\bf E}|<|B|$ we have a vanishing result, since the case is
equivalent to one with a pure magnetic field.

\section{1+1 dimensions}           

In 1+1 dimensions there is no magnetic field, and the electric field
has only one component. Thus the problem is still simpler. As
before, we begin with Eqs. (7) and (11). Consider the electric field
\bb
{\bf E}=E{\bf e}_x,
\ee            
where $E$ is a constant which may be positive or negative. 
The nonvanishing component of $F_{\mu\nu}$ is $F_{01}=E$, and
\bb
\exp\left(-\textstyle{\frac 12}ies\sigma^{\mu\nu}
F_{\mu\nu}\right)
=\exp(eEs\gamma^0\gamma^1).
\ee         
This is exactly the same as Eq. (39). In 1+1 dimensions the $\gamma$
matrices are also $2\times 2$ ones, thus Eq. (40) remains valid.
With Eqs. (40) and (11), we have a result of the same form as Eq.
(41) where $P^2$, say, is of course different in different dimensions.
It is easy to show that
\bb
\langle x|e^{isP^2}|x\rangle={1\over 4\pi s}.
\ee     
As before, we denote
\bb
X^\mu=(X_0, X),\quad P^\mu=(P_0,P_x),
\ee              
and choose
\bb
A^\mu=(A_0,A_x)=(-EX,0),
\ee                 
which corresponds to the field strength (53), and results in
\bb
(P-eA)^2=(P_0+eEX)^2-P_x^2.
\ee        
Using Eq. (23a) we obtain
\bb
e^{is(P-eA)^2}=e^{iP_0P_x/eE}e^{-is(P_x^2-e^2E^2X^2)}
e^{-iP_0P_x/eE}.
\ee                
It is now quite essy to show that
\bb
\langle x|e^{is(P-eA)^2}|x\rangle={|eE|\over 2\pi}
{\rm tr}\,e^{-is(P_x^2-e^2E^2X^2)}.
\ee        
Using Eq. (28a) we have
\bb
\langle x|e^{is(P-eA)^2}|x\rangle={|eE|\over 4\pi\sinh(|eE|s)}.
\ee     
Combining Eqs. (7), (41), (55), and (61) we arrive at
\bb
W(x)={1\over 2\pi}\int_0^\infty ds\;
{e^{-im^2s}\over s}\left[
|eE|\coth(|eE|s)-{1\over s}\right].
\ee            
Using the identity ${\rm Re}W=(W+W^*)/2$ and
making the change of variable $s\to -s$ in $W$ we obtain
\bb
w={1\over 4\pi}\int_{-\infty}^{+\infty} ds\;{e^{im^2s}\over s}
\left[|eE|\coth(|eE|s)-{1\over s}\right].
\ee            
The integrand is regular everywhere in the integration path, thus no
prescription is necessary here. We close the contour of
integration at infinity by a semicircle in the upper half plane and
evaluate the integral by the residue theorem. The result turns out
to be
\bb
w={|eE|\over 2\pi}\sum_{n=1}^\infty{1\over n}
\exp\left(-{n\pi m^2\over |eE|}\right).
\ee                   
The convergence of the series is obvious. This can also be written
in a closed form
$$
w=-{|eE|\over 2\pi}\ln\left[1-
\exp\left(-{\pi m^2\over |eE|}\right)\right].
\eqno(64')$$

\section{Summary and discussions}           

In this paper we calculate the probability of electron-positron
pair creation
in vacuum in external constant uniform electromagnetic
fields in 3+1 and lower dimensions. The results are also applicable
to other charged fermions of spin $1\over 2$. In 3+1 dimensions the
most general result is given  by Eq. (36), where {\bf E} and {\bf B}
may have arbitrary magnitudes and directions. The result for the
ralatively simple case where ${\bf B}\parallel {\bf E}$ is given by
Eq. (32). For a pure magnetic field the probability vanishes, which
means that pair creation cannot occur in this case, as expected
physically. For a pure electric field, our result reduces to that
of Schwinger, Eq. (33). In 2+1 dimensions the magnetic field has
only one component while the eletric field has two.
When $|{\bf E}|>|B|$ the result is given by Eq. (52). For the special
case of a pure electric field this reduces to Eq. (51). When
$|{\bf E}|<|B|$ the result vanishes since the situation is equivalent
to one with a pure magnetic field. In 1+1 dimensions there is no
magnetic field, and the electric field has only one component. The
result is given by Eq. (64), or by Eq. ($64'$) in closed form.

For a pure electric field, the results in the several cases of
different dimensions studied can be written in a unified form
\bb
w_E=(1+\delta_{d3}){|eE|^{(d+1)/2}\over (2\pi)^d}\sum_{n=1}^\infty
{1\over n^{(d+1)/2}}
\exp\left(-{n\pi m^2\over |eE|}\right),
\ee                   
where $d=1,2,3$ is the spatial dimension, and $\delta_{d3}=1$ when
$d=3$ and vanishes in other cases. The additional factor 2 in 3+1
dimensions is due to the double spin states. More specifically, in
3+1 dimensions, with a given momentum there are two
linearly independent solutions of
positive energy  and two of negative energy
to the free Dirac equation, while in 2+1 or 1+1 dimensions there is
only one solution for each energy.

Pair creation of charged scalar particles 
in vacuum in external electromagnetic
fields can also be studied in a similar way. In this case, however,
the probability is reduced by the presence of a magnetic field,
thus we do not discuss the problem in detail.
                                    
\section*{Acknowledgment}

The author is grateful to Professor Guang-jiong Ni for 
useful communications and for encouragement.
This work was supported by the
National Natural Science Foundation of China.

\section*{Note added in proof}

After this paper was accepted for publication, we became aware of
some recent works dealing with the same or relevant problems
by different methods [7-9]. We thank the authors for bringing their
works to our attention.


\end{document}